\documentstyle[preprint,aps]{revtex}

\begin{document}
\title{The effects of \ the nonextensivity on the dimerization process and nematic
ordering}
\author{O. Kayacan\thanks{
e-mail: ozhan.kayacan@bayar.edu.tr}}
\address{Department of Physics, Faculty of Art and Science, Celal Bayar University,\\
Muradiye/Manisa-TURKEY}
\maketitle

\begin{abstract}
\qquad The influence of the dimerization process on the nematic ordering is
investigated by using a nonextensive thermostatistics, namely Tsallis
thermostatistics(TT). A theoretical model taking into account the
dimerization influence on the nematic scalar order parameter has been
recently presented[1]. Our study considers this model and the theoretical
predictions for the nematic order parameter are improved by using TT. 
\newline

\noindent {PACS Number(s):}~ 05.20.-y, 05.70.-a, 61.30.Cz, 61.30. Gd.\newline

\end{abstract}

\newpage

\section{Introduction}

\qquad Since the papers by Tsallis [2,3], Tsallis thermostatistics (TT) has
been commonly used for the investigation of the physical systems.
Boltzmann-Gibbs statistics is a powerfull tool to study a variety of the
physical systems. However it fails for the systems which $i)$ have
long-range interactions, $ii)$ have long-range memory effects, and $iii)$
evolve in multi-fractal space-time. The system which has the above
properties is called ''nonextensive one''. If investigating a nonextensive
system, one then needs an appropriate statistics, and TT is one of these
statistics. It is the aim of this paper to enlarge the applications of TT to
the liquid crystal systems. In doing so, we first summarize the axioms of
TT, and then investigate the influence of the dimerization process on the
nematic ordering by using TT.

TT has been applied to various concepts of thermostatistics and achieved in
solving some physical systems, where Boltzmann-Gibbs statistics is known to
fail [4]. Recently, MST has been generalized within TT and generalized\ MST
has been applied to a nematic liquid crystal, p-azoxyanisole ($PAA$), in [5]
in which Kayacan et al. used the second choice for the internal energy
constraint which will be given below.

TT considers three possible choices for the form of a nonextensive
expectation value. These choices have been studied in [6] and applied to two
systems; the classical harmonic oscillator and the quantum harmonic
oscillator. In that study, Tsallis et al. studied three different
alternatives for the internal energy constraint. The first choice is the
conventional one and used in [2] by Tsallis

\begin{equation}
\sum_{i=1}^{W}p_{i}\varepsilon _{i}=U_{q}^{(1)}.
\end{equation}
The second choice is given by

\begin{equation}
\sum_{i=1}^{W}p_{i}^{q}\varepsilon _{i}=U_{q}^{(2)}
\end{equation}
and regarded as the canonical one. Both of these choices have been applied
to many different systems in the last years [7]. However both of them have
undesirable difficulties. The third choice for the internal energy
constraint is

\begin{equation}
\frac{\sum_{i=1}^{W}p_{i}^{q}\varepsilon _{i}}{\sum_{i=1}^{W}p_{i}^{q}}%
=U_{q}{}^{(3)}.
\end{equation}
This choice is commonly considered to study physical systems because it is
the most appropriate one, and is denoted as the Tsallis-Mendes-Plastino
(TMP) choice. $q$ index is called the entropic index and comes from the
entropy definition [2],

\begin{equation}
S_{q}=k\frac{1-\sum_{i=1}^{W}p_{i}^{q}}{q-1},
\end{equation}
where $k$ is a constant, $\sum_{i}p_{i}=1$ is the probability of the system
in the $i$ microstate, $W$ is the total number of configurations. In the
limit $q\rightarrow 1$, the entropy reduces to the well-known
Boltzmann-Gibbs (Shannon) entropy.

\qquad The optimization of $S_{q}$ leads to

\begin{equation}
p_{i}^{(3)}=\frac{\left[ 1-(1-q)\beta (\varepsilon
_{i}-U_{q}^{(3)})/\sum_{j=1}^{W}(p_{j}^{(3)})^{q}\right] ^{\frac{1}{1-q}}}{%
Z_{q}^{(3)}}
\end{equation}

with

\begin{equation}
Z_{q}^{(3)}=\sum_{i=1}^{W}\left[ 1-(1-q)\beta (\varepsilon
_{i}-U_{q}^{(3)})/\sum_{j=1}^{W}(p_{j}^{(3)})^{q}\right] ^{\frac{1}{1-q}}.
\end{equation}
This equation is an implicit one for the probabilities $p_{i}$. Therefore
the $normalized$ $q-expectation$ $value$ of an observable is defined as

\begin{equation}
A_{q}=\frac{\sum_{i=1}^{W}p_{i}^{q}A_{i}}{\sum_{i=1}^{W}p_{i}^{q}}%
=\left\langle A_{i}\right\rangle _{q}
\end{equation}

where $A$ denotes any observable quantity which commutes with the
Hamiltonian. This expectation value recovers the conventional expectation
one when $q=1$. As mentioned above, Eq.(7) is an implicit one and in order
to solve this equation, Tsallis et al. suggested two different approaches; ''%
$iterative$ $procedure$'' and ''$\beta \rightarrow \beta ^{^{\prime }}$''
transformation.

Since the present study is based on a theoretical model [1], which has been
recently presented, at this stage we give a summary of this model.

The contituent molecule in the liquid crystals is assumed as a rigid rod in
the most commonly used theoretical models. Maier-Saupe theory [8,9] and this
approximation explain the most of the macroscopic observable in nematics.
However, this approximation not appropriate to explain the observed
anomalies in 4,n-alkyloxybenzoic acids [10-12]. They have nematic and
smectic phases. The anomalies are in the temperature dependence of the
elastic constants[13], the electroconductivity [14], and the dielectric
permitivity [15]. The variation of these constants strongly deviate from
those of the classical nematics. The unusual thermal behaviour of the
nematic phase of 4,n-alkyloxybenzoic acids was found and investigated by
some optical methods. The experimental results show a strong textural
transition and an increasing of the depolarized Rayleigh scattering at a
given temperature T$^{\ast }$ in the nematic range [11]. It is reasonable to
assume that this phenomenon is connected to the temperature variable form of
the constituent molecules of the 4,n-alkyloxybenzoic acids. These substances
change the molecular structure with the temperature between three states,
which are closed dimers, open dimers and monomers [16,17]. The variations of
the closed dimer, open dimer and monomer equilibrium concentrations vs
temperature were detected by FT-IR spectroscopy [17]. On the dimer ring,
realized by the pairing of two monomers in a closed dimer by hydrogen
bonding, are imposed different conformational states, which were detected by
far FT-IR spectroscopy [18-19]. A deviation of the temperature trend of the
order parameter in the nematic phase of the NOBA (nonyloxybenzoic acids) was
demonstrated [20]. A similar temperature trend for NOBA was found in [21].
All these experimental results indicate that the dimarization process
strongly influences the nematic ordering in the 4,n-alkyloxybenzoic acids.

Now, we discuss the classical theories. A model keeping the mean field
approximation, but considering the liquid crystal molecules as pairs of
monomers [22] or as association of structural elements in flexible molecules
with internal degrees of freedom [23] were presented. These theories are
motivated by the insufficiency of the classical models, which approximate
the liquid crystal molecules as rigid rods, to explain the phenomena like
reentrant liquid crystal state modeled [22] and the isostructural phase
transitions [23]. Barbero et al. found a similarity in the liquid crystal
state modeled [22] as a mixture of monomers and dimers, used to explain the
provocation of the reentrant behaviour and the nematic state of the
4,n-alkyloxybenzoic acids detected by FT-IR spectroscopy also like mixture
of these molecular forms. However this model can not be used for some
quantitative predictions, since dimerization described in [22] is considered
as a result of dipole-dipole interaction of the polar end groups of the
monomers.

Pershin et al. [23] propose a model which takes into account the influence
of the molecular structure on the macroscopic properties of the nematics,
consisting of flexible molecules like those HOBA (heptyloxybenzoic acids),
OOBA (octyloxybenzoic acids) and NOBA (nonyloxybenzoic acids), and predicts
isostructural transitions in the nematic phase. These transitions already
were detected by DSC and impedance spectroscopy [24,25]. On the other hand,
this theory also can not explain the scalar order parameter vs temperature
and dimer, monomer concentration temperature variations, because the theory
considers the molecular conformational order-disorder in a general way and
the structural element used as a conformational unit is not determined as a
concrete part of the molecule.

Because of these insufficiencies, Barbero et al. presented a theoretical
model [1] and it is an improved one of the classical mean field theories. It
takes into account the internal degree of freedom of the molecules due to
the variation between the three molecular forms: closed dimers, open dimers
and monomers. In their model, the energy of the hydrogen bonding is one of
the quantities connected with the influence of the dimerization process on
the macroscopic properties. This energy was measured in [16]. It is
important to note that Barbero et al. have not taken into account the short
range forces and fluctuations, which leads to a smectic short range order in
a coexistence with the nematic long range order. Our study can be considered
to succeed this disadvantages. So this study is a further development of the
classical theories, and of the theory proposed by Barbero et al.

We use the same notation as Barbero et al.; $N$ is the total number of the
closed dimers, at $T=0$; $N^{\ast }$, $N_{m}=2N^{\ast }$, $N_{c}$ and $N_{o}$
are the total number of dissociated dimers, monomers, closed dimers and open
dimers at a given temperature $T$. $E>0$ is the energy necessary to break a
hydrogen bond of the closed dimer, the activation energy of the chemical
reaction 1 closed dimer$\rightarrow $1 open dimer is then $E$; the
activation energy of the chemical reactions 1 closed dimer$\rightarrow $2
monomers is $2E$ and 1 open dimer$\rightarrow $2 monomers is $E$.

At a given temperature, the equilibrium distributions of open dimers, closed
dimers and monomers are given by: 
\begin{eqnarray*}
n_{o} &=&\exp (-(\mu +E)/k_{B}T) \\
n_{c} &=&\exp (-\mu /k_{B}T) \\
n_{m} &=&2\exp (-(\mu +2E)/k_{B}T)
\end{eqnarray*}

with the condition $n_{o}+n_{c}+n_{m}/2=1$, where $\mu $ is the chemical
potential of the mixture, $n_{o}=N_{o}/N$, $n_{c}=N_{c}/N$, $n_{m}=N_{m}/N$.

Now we consider the nematic order in the mixture, which is formed by closed
dimers, open dimers and monomers. Barbero et al. assumed that monomers did
not contribute to the nematic order. This points out to suppose that the
monomers are nearly of spherical shape. Although this assumption is not too
far from the reality, it could affect the predictions of the model proposed
by Barbero et al. As can be further, our results is in very good agreement
with the experimental data. In their study, the interactions among the
molecules of the closed and open dimers are responsible for the nematic
order. They supposed also that the closed and open dimers were formed by
rod-like molecules. Let ${\bf n}$ be the mean nematic order of the mixture.
The angle between the molecular long axis of the closed dimer and ${\bf n}$
will be indicated $\theta _{c}$, and the angle between that of the open
dimer and ${\bf n}$ will be indicated $\theta _{o}$, Therefore the total
nematic potential is given by 
\begin{equation}
V(\theta _{o},\,\theta _{c})=\sum_{i,j}V_{i,j}=V_{oo}+V_{cc}+V_{oc}+V_{co}
\end{equation}
in the mean field approximation, where $V_{oo}$ and $V_{cc}$ are the mean
field potentials acting on a molecule of open dimer due to the other open
dimers and closed dimers, respectively. $V_{cc}$ and $V_{co}$ have similar
meanings; $V_{cc}$ and $V_{co}$ are the potentials acting on a molecule of
closed dimer due to the other closed dimers and open dimers, respectively.
The partial nematic mean field potentials $V_{i,j}$ are given by [26] 
\begin{equation}
V_{i,j}=-\alpha _{i,j}P_{2}(\theta _{i})\left\langle P_{2}(\theta
_{j})\right\rangle
\end{equation}
in the Maier-Saupe approximation, where the coupling constants $\alpha
_{i,j} $ depend on the distance between the centers of mass of the molecules
and on a molecular property.

Eq.(8) can be rewritten by using Eq.(9) as the following: 
\begin{equation}
V(\theta _{o},\theta _{c})=V_{o}(\theta _{o})+V(\theta _{c})
\end{equation}
where 
\begin{equation}
V_{i}(\theta _{i})=-\left[ \alpha _{ii}S_{i}+\alpha _{ij}S_{j}\right]
P_{2}(\theta _{i}).
\end{equation}
In Eq.(11), we use the relation $S_{i}=\left\langle P_{2}(\theta
_{i})\right\rangle $ for the scalar order parameters of closed ($S_{c}$) and
open ($S_{o}$) dimers, respectively, where $P_{2}(\theta _{i})$ is the
second Legendre polynomial. Then the partition function can be given by 
\begin{equation}
f=\int_{0}^{1}d(\cos \theta _{o})\int_{0}^{1}d(\cos \theta _{c})\exp
(-(V_{o}+V_{c})/k_{B}T)
\end{equation}
and the free energy can be written as 
\begin{eqnarray}
F &=&U-TS_{e}  \nonumber \\
&=&-Nk_{B}T\;\ln f+\frac{1}{2}N\sum_{i,j}\alpha _{i,j}S_{i}S_{j},
\end{eqnarray}
where $S_{e}$ is the entropy of the system. The scalar order parameters are
determined from the following relation: 
\begin{equation}
S_{i}=\frac{\int_{0}^{1}d(\cos \theta _{i})P_{2}(\theta _{i})\exp (-\beta
V_{i})}{f_{i}},
\end{equation}
where $i$ stands for $o,c$; open dimers, closed dimers, $\beta =1/(k_{B}T)$.
Eq.(14) is a self-consistent equation and can be solved numerically, and
also depends on; $a)$ the temperature $T$, $b)$ the activation energy $E$,
and $c)$ the quantities $\alpha _{i,j}$ which contains the strengths of the
interaction closed dimer-closed dimer, closed dimer-open dimer and open
dimer-open dimer. Barbero et al. made an analysis and assumed that the
concentration dependence of these quantities was determined by 
\begin{equation}
\alpha _{i,j}=\alpha _{i,j}^{n}=u_{i,j}\;n_{j}^{2}.
\end{equation}
We use the same assumption in this study. The nematic order parameters of
the closed ($S_{c}$) and open ($S_{o}$) dimers can be determined from
Eq.(14). If $S_{o}$ and $S_{c}$ are known, then the nematic order parameter
of the mixture is calculated by using the following relation: 
\begin{equation}
S=\frac{n_{c}\;S_{c}+n_{o}\;S_{o}}{n_{c}+n_{o}}.
\end{equation}
Let us now discuss how the order parameter of the mixture is calculated. The
partition function of the mixture is written as 
\begin{eqnarray}
f_{q} &=&\int_{0}^{1}\int_{0}^{1}d(\cos \theta _{o})\;d(\cos \theta
_{c})\exp _{q}(-(V_{o}+V_{c})/k_{B}T  \nonumber \\
&=&\int_{0}^{1}\int_{0}^{1}d(\cos \theta _{o})\;d(\cos \theta _{c})\;\left[
1+(1-q)(V_{o}+V_{c})/k_{B}T\right] ^{\frac{1}{1-q}}
\end{eqnarray}
within TT. The order parameters are evaluated from Eq.(7), and this equation
is written in TT formalism as: 
\begin{equation}
S_{i,q}=\frac{\int_{0}^{1}(d\cos \theta _{i})\;p_{i}^{q}S_{i}}{%
\int_{0}^{1}(d\cos \theta _{i})\;p_{i}^{q}}=\left\langle S_{i}\right\rangle
_{q},
\end{equation}
where $i=o,c$. If the partial order parameters are evaluated from this
equation, the nematic scalar order parameter of the mixture is calculated
from Eq.(16). The critical value of the order parameter, in equilibrium, is
determined from Eq.(13).

\section{Results and Discussion}

The nematic-isotropic transition being first-order, occurs when the
orientational Helmholtz free energy vanishes, provided the volume change is
negleted, and so to investigate this transition the free energy is
calculated from the interaction potential energy.

In the Fig.(1), the nematic order parameter of the mixture $S$ vs
temperature is shown. $q=1$ curve represents the Barbero et al. model. It is
clear that the curve for $q=1.013$ which is plotted by assuming $\alpha
_{i,j}\propto \;n_{j}^{2}$ , $u_{cc}\simeq 1\;eV$ [26], $%
u_{oo}=u_{co}=u_{oc}=u_{cc}$, and $E=4.8kcal/mole$ [16] is in very good
agreement with the experimental data. Fig.(2) shows the free energy vs the
order parameter of the mixture at the phase transition temperature. The
curves have two minima; one at $s=0$ and the other at $s=s_{c}.$ $s=0$
corresponds to the liquid phase, and $s\neq 0$ corresponds to the nematic
phase. The minimum corresponding to the nematic phase for $q=1$ occurs at $%
s_{c}\cong 0.429$. Since The Barbero et al. model is based on the
Maier-Saupe theory, this is not a surprising result. It is well known that
Maier-Saupe theory gives a universal value,$\;0.429$, of the order parameter
at the nematic-isotropic transition temperature. However, the minimum
corresponding to the nematic phase for $q=1.013$ occurs at $s_{c}\cong 0.442$%
. As can be seen from these results, the minimum of the free energy,
corresponding to the nematic phase, changes to a higher value as $q$ is
increased. The experimental results indicate that the critical value of the
scalar order parameter in nematic-isotropic transition has a value between $%
0.25-0.5$ [27]. Although the theory based on the Maier-Saupe theory does
predict the variation of the order parameter, the quantitive predictions of
the theory which yields an interaction potential with just one parameter can
not account precisely for such properties as the temperature dependence of
the orientational order $\left\langle S_{i}\right\rangle $ [28]. Then
generalized form of the standard theory could give a different value of the
order parameter from $0.429,e.g.$ $0.442$. When this generalized form of the
theory based on the Maier-Saupe theory is used (e.g. inserting another
parameter, $q$ (entropic index) coming from the TT), the critical value of
the order parameter appears to assume a value lying in this range and the
temperature dependence of the order parameter is explained in very good
agreement with the experimental results. As a consequence, with a small
departure from the standard theory, one could explain the behaviour of the
order parameter vs temperature and the nematic-isotropic phase transition.

As becomes clear from the results obtained by employing TT, the present
study can also be applied to other liquid crystal systems. In this context,
a theoretical model, which is proposed to investigate the influence of the
dimerization process on the nematic ordering, has been employed and the
theoretical results have been improved by using TT.

\section*{References}

[1] G. Barbero, L.R. Evangelista, M.P. Petrov, Phys. Lett. A 256 (1999) 309.

[2] C. Tsallis, J.Stat. Phys. 52 (1988) 479.

[3] E.M.F. Curado, C. Tsallis, J. Phys. A 24 (1991) L69.

[4] S. Abe, Y. Okamoto, Nonextensive Statistical Mechanics and Its
Applications, Series Lectures Notes in Physics, Berlin: Springer-Verlag,
2001.

[5] O. Kayacan, F.B\"{u}y\"{u}kk\i l\i \c{c}, D. Demirhan, Physica A 301
(2001) 255.

[6] C. Tsallis, R.S. Mendes, A.R. Plastino, Physica A 261 (1998) 534.

[7] see http://tsallis.cat.cbpf.br for an updated bibliography.

[8] W. Maier, A. Saupe, Z. Naturforsh. 14a (1959) 889; 15a (1960) 287.

[9] P.G. de Gennes, The Physics of Liquid Crystals, Oxford Univ. Press,
Oxford, 1974.

[10] P.D. Simova, M.P. Petrov, Phys. Status Solidi A 80 (1983) K153.

[11] M.P. Petrov, P.D. Simova, J.Phys. D Appl. Phys. 18 (1985) 293.

[12] M.P. Petrov, A. Braslau, A.M. Levelut, G. Durand, J. Phys. II (France)
2 (1992) 1159.

[13] H. Gruler, G. Meier, Mol. Cryst. Liq. Cryst. 23 (1973) 261.

[14] F. Rondelez,, Sol. State Commun. 11 (1972) 1675.

[15] E.F. Carr, Phys. Rev. A 12 (1975) 327.

[16] B. Deloche, B. Cabane, Mol. Cryst. Liq. Cryst. 19 (1972) 25.

[17] M.P. Petrov, E. Anachkova, N. Kirov, H. Ratajczak, J. Baran, J. Mol.
Liq. 61 (1994) 221.

[18] K. Antonova, M.P. Petrov, N. Kirov, H. T. Tenev, H. Ratajczak, J.
Baran, J. Mol. Struct. 325 (1994) 189.

[19] M.P. Petrov, K. Antonova, N. Kirov, T. Tenev, H. Ratajczak, J. Mol.
Struct. 327 (1994) 265.

[20] M.P. Petrov, K. Antonova, N. Kirov, H. Ratajczak, J. Baran, J. Mol.
Struct. 444 (1998) 91.

[21] A.S. Rao, P.N. Murthy, C.R.K. Murthy, T.R.S. Reddy, Phys. Status Solidi
A 68 (1981) 373.

[22] L. Longa, W.H. de Jeu, Phys. Rev. A 26 (1982) 1632.

[23] K. Pershin, V.A. Konoplev, Liq. Cryst. 12 (1992) 95.

[24] S. Frunza, L. Frunza, A. Sparavigna, M.P. Petrov, S.I. Torgova, Mol.
Mat. 6 (1996) 215.

[25] B. Montrucchio, A. Sparavigna, A. Strigazzi, Liq. Cryst. 24 (1998) 841.

[26] G. Vertogen, W.H. de Jeu, Thermotropic Liquid Crystals, Fundamentals,
Springer, Berlin, 1988.

[27]. A. Beguin, J.C. Dubois, P. Le Barny, J. Billard, F. Bonamy, J.M.
Busisine, P. Cuvelier, Sources of thermodynamics data on mesogens, Mol.
Cryst. Liq. Cryst. 115 (1984) 1.

[28] R.L. Humphries, P.G. James, G.R. Luckhurst, J. Chem. Soc. Faraday
Trans. II 68 (1972) 1031.

\newpage

{\bf Figure Caption} \newline

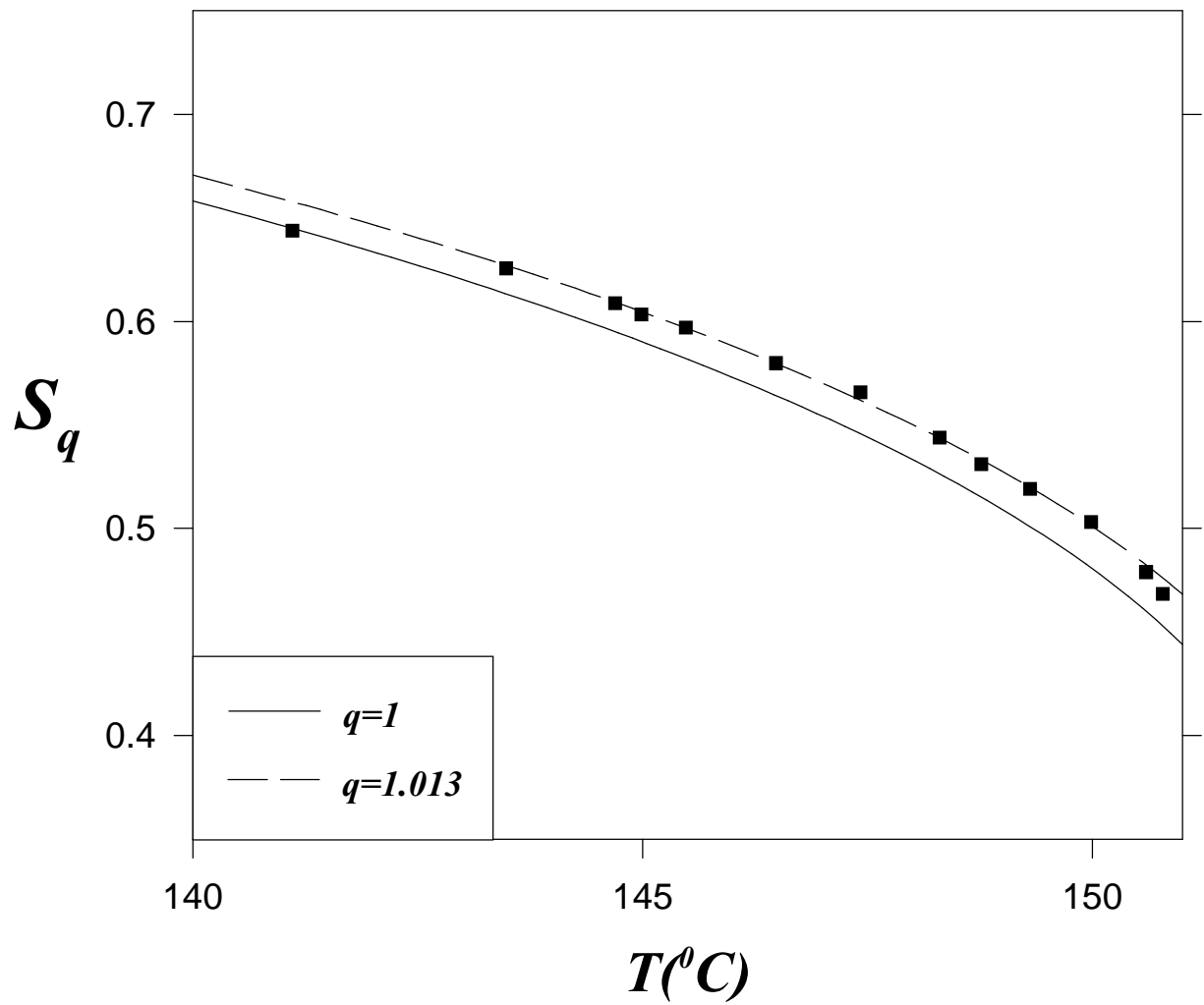
Figure 1. Orientational order parameter as a function of temperature in the
mixture for $q=1.013$ and $q=1$. Filled squares represent the experimental
data[16].\newline

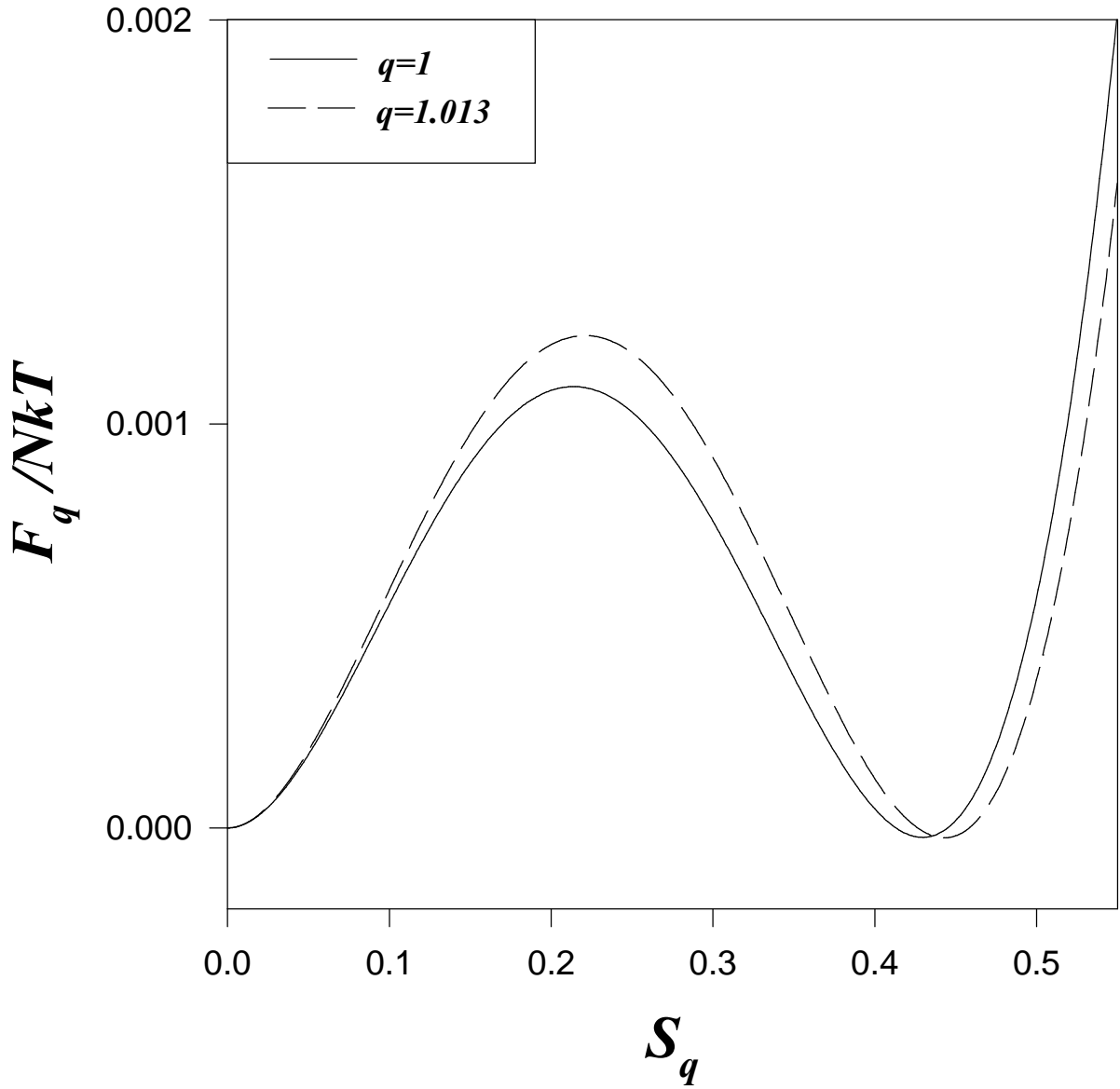
Figure 2. The Helmholtz free energy as a function of generalized order
parameter in the mixture for $q=1.013$ and $q=1$.\newline

\end{document}